\documentclass[useAMS,usenatbib,usegraphicx]{mn2e}
\usepackage{psfig}   
\usepackage{graphicx}
\usepackage{amssymb}
\usepackage{subfigure}

\title[A triggered Solar System?]{Did the Solar System form in a sequential triggered star formation event?}

\author[R. J. Parker \& J. E. Dale]{Richard J. Parker$^{1}$\thanks{E-mail: R.J.Parker@ljmu.ac.uk} and James E. Dale$^{2,3}$  \vspace*{0.1cm} \\
$^{1}$Astrophysics Research Institute, Liverpool John Moores University, 146 Brownlow Hill, Liverpool, L3 5RF, UK\\
$^{2}$Excellence Cluster `Universe', Boltzmannstra{\ss}e 2, 85748 Garching, Germany\\
$^{3}$Universit{\"a}ts-Sternwarte M{\"u}nchen, Scheinerstra{\ss}e 1, 81679 M{\"u}nchen, Germany
}

\begin{document}

\date{}
                             
\pagerange{\pageref{firstpage}--\pageref{lastpage}} \pubyear{2015}

\maketitle

\label{firstpage}

\begin{abstract}
The presence and abundance of the short-lived radioisotopes (SLRs) $^{26}$Al and $^{60}$Fe during the formation of the Solar System is difficult to explain unless the Sun formed in the vicinity of one or more massive star(s) that exploded as supernovae. Two different scenarios have been proposed to explain the delivery of SLRs to the protosolar nebula: (i) direct pollution of the protosolar disc by supernova ejecta and (ii) the formation of the Sun in a sequential star formation event in which supernovae shockwaves trigger further star formation which is enriched in SLRs.

The sequentially triggered model has been suggested as being more astrophysically likely than the direct pollution scenario. In this paper we investigate this claim by analysing a combination of $N$-body and SPH simulations of star formation. We find that sequential star formation would result in large age spreads (or even bi-modal age distributions for spatially coincident events) due to the dynamical relaxation of the first star-formation event(s). Secondly, we discuss the probability of triggering spatially and temporally discrete populations of stars and find this to be only possible in very contrived situations. Taken together, these results suggest that the formation of the Solar System in a triggered star formation event is as improbable, if not more so, than the direct  pollution of the protosolar disc by a supernova.
\end{abstract}

\begin{keywords}   
stars: formation -- kinematics and dynamics -- open clusters and associations: general -- methods: numerical -- planets and satellites: formation -- ISM: H{\small II} regions
\end{keywords}

\section{Introduction}

One of the outstanding challenges in star and planet formation is to understand and characterise the birth environment of our Solar System \citep[e.g.][]{Adams10,Pfalzner15b}. One potential constraint is the presence of the daughter-isotopes of short-lived radioisotopes (SLRs) found in meteorites thought to originate from the epoch of planet formation around the Sun \citep{Lee76}.  Their abundances and short half-lives suggest a rapid inclusion in meteorites during the early stages of Solar System formation.

Several radioisotopes with half-lives ranging from tens of days to several Myr are present in meteorites, but two -- $^{26}$Al and $^{60}$Fe -- were most probably produced by nucleosynthesis in the cores of massive ($>$20\,M$_\odot$) stars \citep{Goswami04}. It is possible to produce $^{26}$Al by cosmic ray spallation \citep{Lee98,Shu01} and both $^{26}$Al and $^{60}$Fe  from pollution from Asymtotic Giant Branch (AGB) stars \citep{Busso99}. However, AGB stars are rare in young star-forming regions \citep{Kastner94} and spallation alone cannot be responsible for the enrichment due to the absence of  $^{60}$Fe. For this reason, the supernovae of massive stars are thought to be the most likely origin of these isotopes. The exact details of the delivery mechanism(s) for these isotopes are still debated, and there are two main hypotheses.

In the first, referred to as ``disc pollution'', $^{26}$Al and $^{60}$Fe are delivered directly to the Sun's protoplanetary disc from the supernova of one or more massive stars. \citet{Chevalier00} and \citet{Ouellette07} estimate that in order to receive the required amounts of enrichment, the Sun must have been between $\sim$0.1 -- 0.3\,pc from the supernova so as to strike a balance between capturing enough of the ejecta without destroying too much of the disc \citep[e.g.][]{Armitage00,Scally01,Adams04}. 

At first glance, the disc pollution scenario  appears rather improbable. \citet{Parker14a} use $N$-body simulations of star-forming regions to determine the fraction of Sun-like stars that are enriched by a supernova explosion at distances $\sim$0.1 -- 0.3\,pc from massive stars when they explode, $f_{\rm enrich}$. They then considered whether each enriched Sun-like star had always been a `singleton' \citep{Malmberg07b} -- i.e. never in a binary star system, $f_{\rm enrich, sing.}$, and finally, whether these enriched singletons had suffered perturbing encounters  with other stars in this dense stellar environment, $f_{\rm enrich, sing., unp.}$. 

The fraction of enriched, single, unperturbed stars is of order $f_{\rm enrich, sing., unp.} \sim 1$\,per cent \citep{Parker14a} -- see also \citet*{Adams14} -- and is calculated by summing together 20 similar simulations (to improve statistics) and assuming that the star-forming regions all disperse into the Galactic field. If one determines this fraction on a region-by-region basis, then some simulations eject the supernova progenitor before enrichment \citep[due to multiple interactions with other massive stars, e.g.][]{Allison11,Oh15} and so $f_{\rm enrich, sing., unp.}$ can often be zero \citep[see also][]{Gounelle08,Adams14}.

Alternatively, several authors have proposed that the delivery of  $^{26}$Al and $^{60}$Fe comes from the giant molecular cloud (GMC) from which the Sun formed. This second scenario, which we will refer to as ``sequential triggering'' \citep{Gaidos09,Gounelle09,Gounelle12,Gritschneder12,Gounelle15}, postulates that $^{26}$Al and $^{60}$Fe were delivered to the GMC in a series of star-forming events, and the supernova that delivered  $^{60}$Fe triggered the formation of the Sun. 

\citet{Gounelle12} provide a detailed picture of this scenario. A first star formation event of several thousand stars contains several massive stars whose supernovae enrich the nearby GMC and trigger a second star formation event, 10\,Myr after the first (in order to allow some of the $^{60}$Fe to decay). This second star formation event also contains $\sim$1000 stars to increase the probability of that event containing a Wolf-Rayet (WR) star. WR winds are rich in $^{26}$Al, and in the \citet{Gounelle12} model the wind of this WR star causes a third generation of stars to form. It is in this third generation that the Sun -- with the correct levels of $^{26}$Al and $^{60}$Fe -- forms. 

The sequential triggering scenario has been suggested as a way of enriching the Solar System meteorites without the need for the improbable astrophysical conditions required by the disc pollution scenario (close but non-destructive encounter with supernova, no previous or subsequent interactions with the $\sim$2000 -- 4000 other stars in the region, efficient coupling of hot supernova ejecta to the cold disc material). 

In principle,  $^{26}$Al and $^{60}$Fe could also be delivered to the Solar System if stars form spontaneously (as opposed to triggering) in a GMC that has already been polluted. We briefly comment on this scenario in Sections~\ref{trigger}~and~\ref{conclude}.

In this paper we investigate whether the sequential triggering scenario is more likely than the disc pollution scenario. We start by discussing the dynamical implications of  multiple star formation events through triggering in Section~\ref{dynamics}. We then discuss the efficiency and nature of triggered star formation in hydrodynamical simulations in Section~\ref{trigger}. We conclude in Section~\ref{conclude}.

\section{Dynamical evolution of sequential events}
\label{dynamics}

The triggered sequential star formation scenario suggested by \citet{Gounelle12} requires three discrete star formation events, each forming $\sim$1000 stars, with the first two separated in time by roughly 10\,Myr and a third occuring $\sim$15\,Myr after the first.   

If several star-forming events did sequentially lead to the formation of the Solar System, then these events must remain largely discrete, due to the absence of observed age spreads of more than several Myr in nearby star-forming regions \citep{Reggiani11b,Jeffries11}. The triggered scenario for the formation of the Sco Cen OB association discussed in \citet{Preibisch07} suggests that the sequential events must be spatially discrete due to  a lack of age spreads within the sub-groups.

Star-forming regions do not remain static, however. Multiple simulations \citep[e.g.][]{Klessen01,Moeckel10,Gieles12,Parker12d,Parker13a} show that two-body interactions cause a region to relax by expanding, so that the median stellar density may decrease by several orders of magnitude within 10\,Myr. This implies a large amount of expansion, which means that if supernovae and other feedback mechanisms are triggering further star formation events, they must remain spatially discrete even after dynamical evolution.

In addition to two-body relaxation, the expulsion of residual gas has been shown to cause the expansion of  star-forming regions due to a rapid decrease in the total potential \citep[e.g.][and many more]{Tutukov78,Hills80,Lada84,Goodwin06,Baumgardt07,Pfalzner14}, if the gas is treated as a background potential. Simulations which include a more realistic treatment of the gas \citep{Offner09,Kruijssen12a,Dale12b} suggest that when the local star formation efficiency is high, removing this remnant gas does not lead to significant expansion of the regions. The principal reason is that the local star formation efficiency within the subclusters is high, and so the potential there is dominated by the stars. Removing any leftover gas therefore has a negligible effect on the potential inside the subclusters, or on their subsequent dynamical evolution in these simulations. The contribution (if any) of gas expulsion to the expansion of star-forming regions is, however, still hotly debated in the literature.

Regardless of the agent of expansion, if sequential triggered star formation did form the Solar System, a minimum distance between events must be required to prevent significant, or even bi-modal age spreads of 10 -- 15\,Myr  which would presumably be observable in similar star-forming regions today \citep{Preibisch07}.  In order to determine this minimum distance, we take a simulation of star-formation from the suite by \citet{Dale14} -- `Run~J', which forms 564 sink-particles, of which more than ten are O-type stars with mass $>$20\,M$_\odot$ and will contribute $^{60}$Fe to the GMC from their supernovae. We then take the sink particle distribution (masses, positions and velocities) in this simulation and evolve it with the \texttt{kira} integrator in the Starlab environment \citep{Zwart99,Zwart01}. 

\begin{figure*}
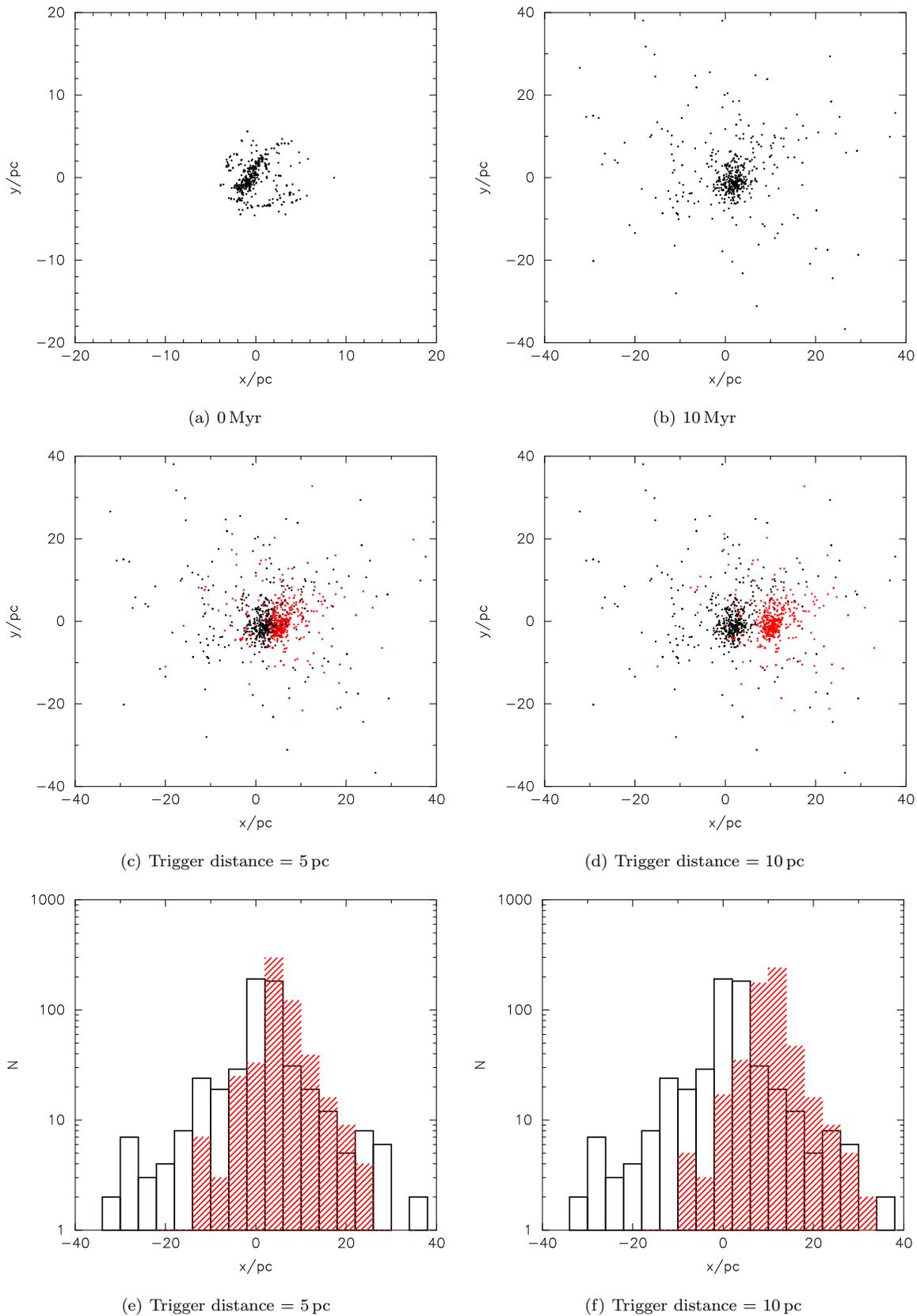

  \begin{center}
\setlength{\subfigcapskip}{10pt}
\hspace*{-1.5cm}\subfigure[0\,Myr]{\label{bg-a}\rotatebox{270}{\includegraphics[scale=0.35]{Dale_RunJ_feedback_0Myr.ps}}} 
\hspace*{0.3cm} 
\subfigure[10\,Myr]{\label{bg-b}\rotatebox{270}{\includegraphics[scale=0.35]{Dale_RunJ_feedback_old_only.ps}}}
\hspace*{-1.5cm}\subfigure[Trigger distance = 5\,pc]{\label{bg-c}\rotatebox{270}{\includegraphics[scale=0.35]{Dale_RunJ_feedback_intermingle_5pc.ps}}} 
\hspace*{0.3cm} 
\subfigure[Trigger distance = 10\,pc]{\label{bg-d}\rotatebox{270}{\includegraphics[scale=0.35]{Dale_RunJ_feedback_intermingle_10pc.ps}}}
\hspace*{-1.5cm}\subfigure[Trigger distance = 5\,pc]{\label{bg-e}\rotatebox{270}{\includegraphics[scale=0.35]{Dale_RunJ_feedback_intermingle_5pc_hist.ps}}} 
\hspace*{0.3cm} 
\subfigure[Trigger distance = 10\,pc]{\label{bg-f}\rotatebox{270}{\includegraphics[scale=0.35]{Dale_RunJ_feedback_intermingle_10pc_hist.ps}}}
\caption[bf]{Dynamical evolution of simulation Run J with dual feedback from \citet*{Parker15a}  at (a) 0\,Myr and (b) 10\,Myr. Panels (c) and (d)  show the simulation at 10\,Myr but also include the same simulation superimposed at an age of 5\,Myr (red points), assuming it represents a second star formation event triggered at a distance of (c) 5\,pc and (d) 10\,pc. Panels (e) and (f) show the numbers of stars along the x-axis for the two superimposed star-formation events; the older event is shown by the black open histogram, and the younger event is shown by the red hashed histogram, assuming it was triggered at a distance of (c) 5\,pc and (d) 10\,pc. }
\label{bg}
  \end{center}
\end{figure*}

Full details of the dynamical evolution of this simulation are described in \citet*{Parker15a}. In Fig.~\ref{bg-a} we show the spatial distribution of the simulation after the SPH calculation described in \citet{Dale14}, but before any subsequent $N$-body evolution.  Fig.~\ref{bg-b} shows the simulation following 10\,Myr of evolution, after the star-formation event has dynamically relaxed and expanded \citep[the median stellar density in this simulation decreases from 51\,stars\,pc$^{-2}$ to 2.5\,stars\,pc$^{-2}$ in 10\,Myr,][]{Parker15a}.

In order to mimic the triggering of a second star formation event, and its subsequent dynamical evolution, we superimpose a second version of the same simulation, but with an age of 5\,Myr, in the same field of view (the red points)\footnote{One caveat here is that we assume that the second star-formation event evolves in a dynamically similar fashion to the first (i.e. it has comparable density). Whilst dynamical evolution is often inherently stochastic, with stark differences in the evolution of simulations with identical initial conditions, all star-forming regions eventually expand due to two-body relaxation.}. We assume that this second event was triggered at a distance of either 5\,pc (Fig.~\ref{bg-c}) or 10\,pc (Fig.~\ref{bg-d}). Clearly, at either distance a significant  fraction of younger stars appear in the line of sight of older stars. In order to quantify this somewhat, we bin the two populations along the x-axis and the resultant histograms are shown in Figs.~\ref{bg-e}~and~\ref{bg-f}. If the second event is triggered at a distance of 5\,pc, then 50\,per cent of stars along the line-of-sight would be observed to have an age 5\,Myr younger than the older population.

\begin{figure*}
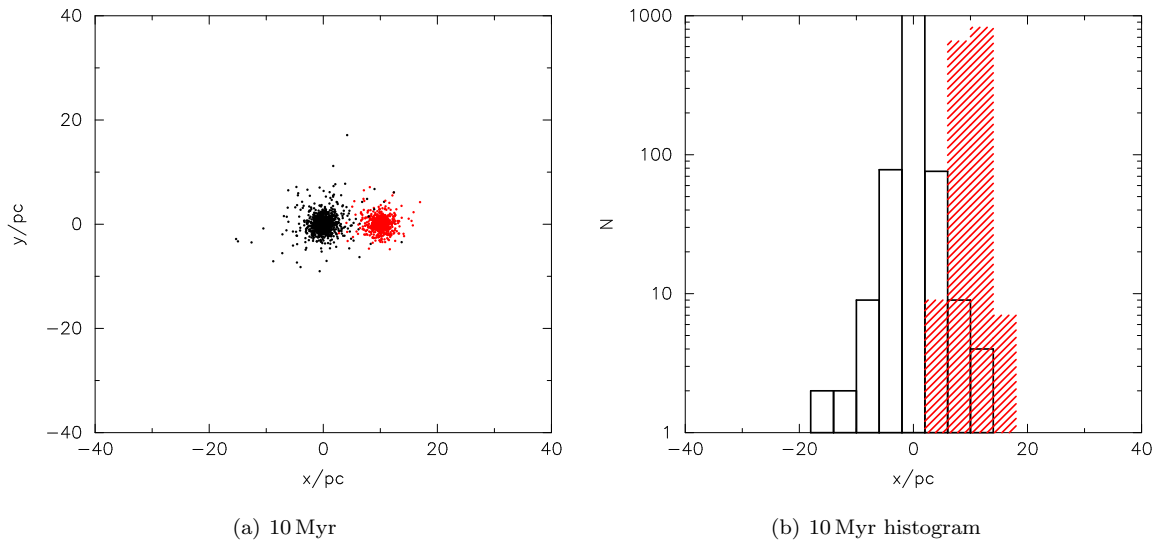

  \begin{center}
\setlength{\subfigcapskip}{10pt}
\hspace*{-1.5cm}\subfigure[10\,Myr]{\label{plummer-a}\rotatebox{270}{\includegraphics[scale=0.35]{plummer_02_intermingle_10pc.ps}}} 
\hspace*{0.3cm} 
\subfigure[10\,Myr histogram]{\label{plummer-b}\rotatebox{270}{\includegraphics[scale=0.35]{plummer_02_intermingle_10pc_hist.ps}}}
\caption[bf]{Dynamical evolution of a simulation of a Plummer sphere after 10\,Myr (black points) and a second triggered event after 5\,Myr (red points), assuming a triggering distance of 10\,pc. In panel (b) we show the numbers of stars along the x-axis for the two superimposed star-formation events; the older event is shown by the black open histogram, and the younger event is shown by the red hashed histogram, assuming it was triggered at a distance of 10\,pc. }
\label{plummer}
  \end{center}
\end{figure*}

So far, we have only considered one simulation, which is rather low density initially and contains only 564 stars. Whilst the low-density reduces the number of interactions that cause the region to expand, the low number of stars more readily leads to the dissoloution of the region \citep{Gieles12,Moeckel12,Parker12d,Parker13a}. If we consider the very dense simulations (initial surface densities of $\Sigma = 10^4$stars\,pc$^{-2}$) presented in \citet{Parker14b} of regions containing $N = 1500$ stars undergoing cool-collapse, we obtain a very similar histogram. The reason is that more stars are ejected due to the high initial density and subsequent violent relaxation, leading to a wide spatial distribution.

We also consider a more benign environment, in which we simulate $N = 1500$ stars in a Plummer sphere \citep{Plummer11} with initial half-mass radius 0.8\,pc (corresponding to a median surface density of $\Sigma = 100$\,stars\,pc$^{-2}$). A Plummer sphere is already a relaxed system and as such is an unrealistic model for the spatial and kinematic outcome of star formation. However, because of its modest dynamical evolution, it does not expand at the same rate as the simulations discussed above (see Fig.~\ref{plummer-a}). When using Plummer spheres to approximate the star-formation events, we see that an observed age bi-modality at a distance of 10\,pc would still be present, but at a much lower level than in the more realistic models of the evolution of star formation environments (Fig.~\ref{plummer-b}).

Based on these simple dynamical models, we suggest that the triggered sequence of star-formation events that formed the Solar System would need to be separated by at least 10\,pc.  Assuming that star-formation environments are initially unrelaxed systems, as observed \citep{Peretto06,Andre10} and corroborated by simulations \citep{Bate12,Dale12b,Dale14,Parker14b}, then in order to avoid age spreads/bi-modality, which are not observed, a distance of at least 40\,pc between events may be required to provide spatially and temporally discrete events.

This is quite a stringent requirement. \cite{Heyer09} analysed a sample of 158 Milky Way molecular clouds, determining their masses and radii. Of these, 139 (88\,per cent) have radii less than 40\,pc, making it unlikely that a star formation event could trigger a second \emph{in the same cloud} at a separation in excess of 40\,pc, and 87 (55\,per cent) have radii less than 20\,pc, making this outright impossible. Such a large required separation likely demands that the triggered population and the triggering agent are formed in different molecular clouds. 

Furthermore, the UV radiation flux from a star-forming region containing $\sim$ 1000 stars with radius 1\,pc is likely to be of order $F_{\rm UV} = 3$\,erg\,s$^{-1}$\,cm$^{-2}$ \citep{Fatuzzo08}. At a distance of 40\,pc from the region this value will decrease to $F_{\rm UV} = 1.8 \times 10^{-3}$\,erg\,s$^{-1}$\,cm$^{-2}$, which is only marginally higher than the ambient value for the interstellar medium \citep[$F_{\rm UV} = 1.6 \times 10^{-3}$\,erg\,s$^{-1}$\,cm$^{-2}$,][]{Habing68}. We will discuss the efficiency of triggering at such distances in the following section.

Finally, we note that 40\,pc is likely to be the minimum \emph{projected} distance that two star formation events could be separated by. In our plots showing two clusters of different ages, the second (triggered) cluster has been superimposed in the field perpendicular to the line-of-sight. In reality, the triggered cluster could appear in the fore- or background of the first cluster, thereby increasing the amount of temporal mixing in the field of view.



\section{Constraints from hydrodynamic simulations}
\label{trigger}

The effects of feedback from massive stars on the star formation process within turbulent molecular clouds, and in particular the ability of \emph{photoionising} feedback to trigger star formation, has recently been investigated by \cite{Dale13b}. In general, it was found that triggering by expanding H{\small II} regions over a timescale of 1-3\,Myr was of rather minor importance, with the principal outcome of photoionising feedback being that \emph{smaller} quantities of stellar mass (and sometimes, fewer stars) were produced in clouds suffering the effects of feedback. In some simulations, significant numbers of stars were caused to form by feedback, but the effect on the total stellar mass and number of stars was offset by disruption of accretion flows onto clusters, and destruction of dense star--forming gas, preventing the formation of stars that would have been born in the absence of feedback.

By taking advantage of the Lagrangian nature of SPH, \cite{Dale13b} identified which stars in their simulations were triggered, based on the criterion of whether or not the gas from which a given object formed was involved in star formation or not in a companion calculation where ionisation was absent. In simulations where sink particles could be treated as individual stars, they found that the number fractions of triggered objects ranged from 10--40\,per cent, and that the mass fractions of triggered objects were similar at 8--37\,per cent. It is therefore possible in principle for a stellar population to trigger the formation of another one of comparable mass, as is required by the model of \citet{Gounelle12}.

\begin{figure*}
  \begin{center}
\setlength{\subfigcapskip}{10pt}
\subfigure[Run I]{\includegraphics[width=0.45\textwidth]{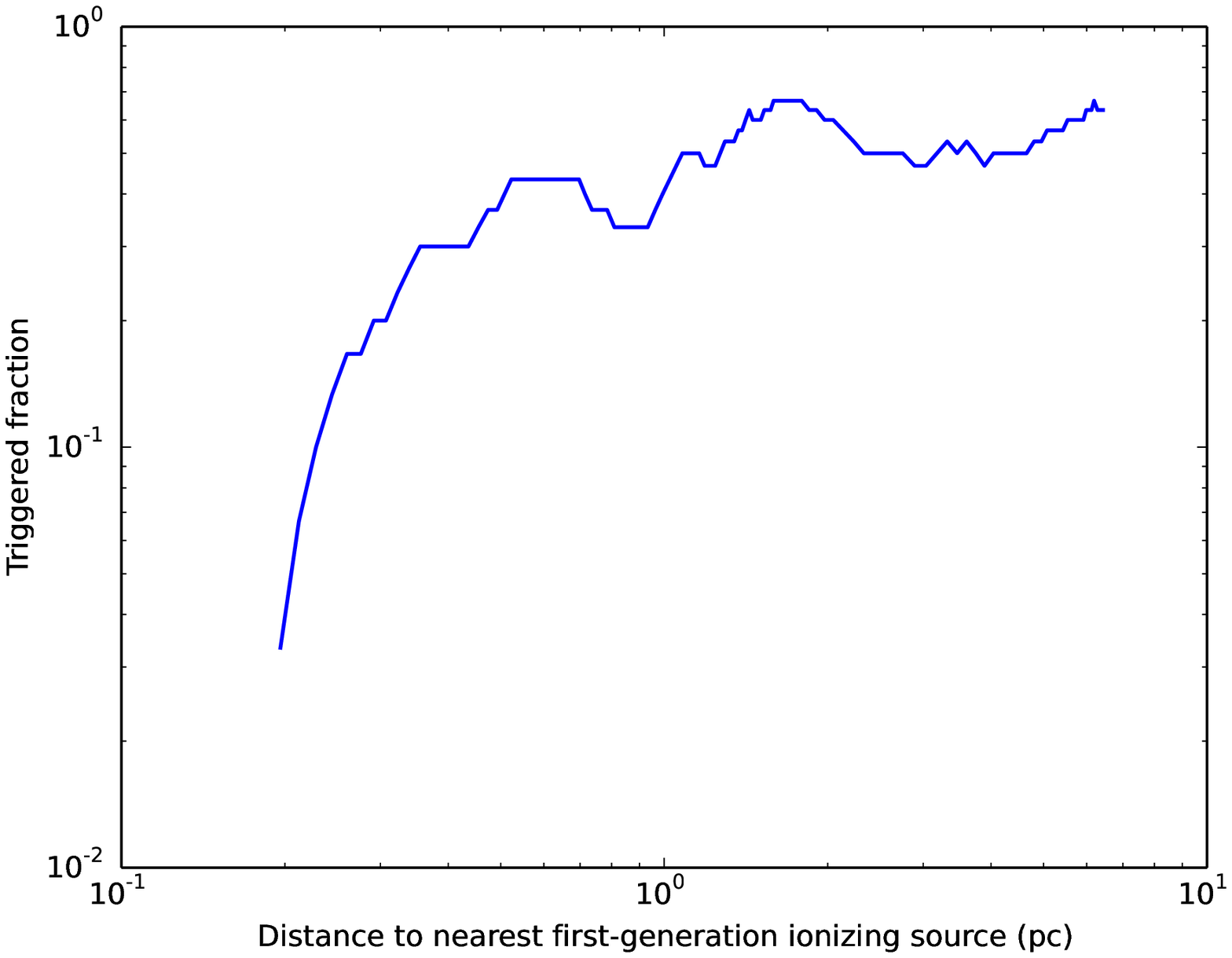}}
\hspace*{1cm} 
\subfigure[Run J]{\includegraphics[scale=0.43]{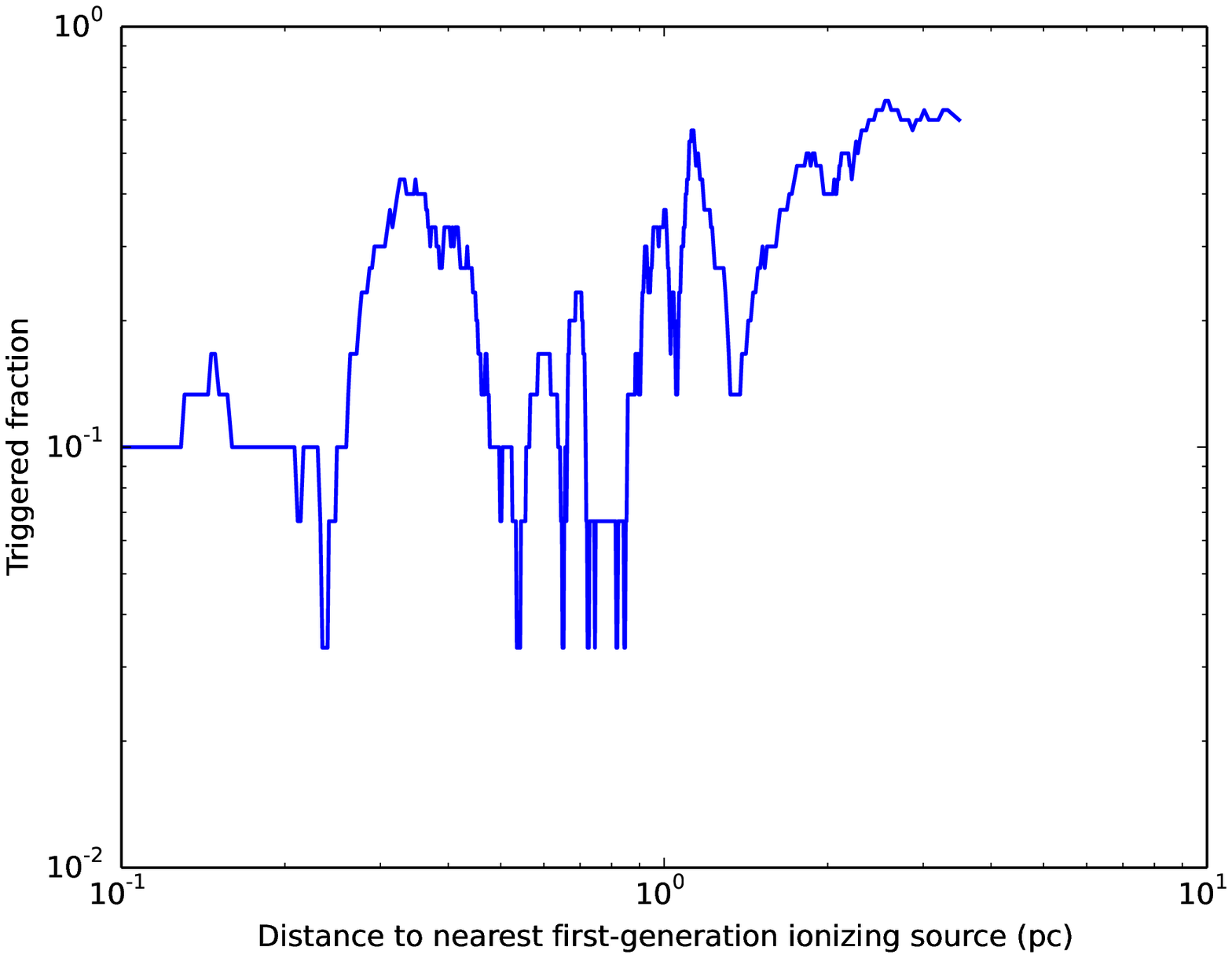}}
\caption[bf]{Fraction of stars which are triggered as a function of three--dimensional distance from the nearest ionising source in Runs I (left panel) and J (right panel) from \cite{Dale13b}.}
\label{fig:trig_vs_r}
  \end{center}
\end{figure*}

However, the main result of the \cite{Dale13b} studies was that the triggered stars were spatially, temporally and dynamically mixed with their spontaneously--formed counterparts. In particular, there was no bimodality observed either in the spatial distribution of the stars (triggered or otherwise) with respect to the ionising stars, or in the ages of objects, whether measured from their time of formation or the point where they acquired their final masses. In some simulations, the fraction of triggered stars increases with three--dimensional distance from the O--type stars, but the increase is not monotonic or sudden, and fails to exceed $\approx60$\,per cent even at the largest distances from the ionising stars, as shown in Figure \ref{fig:trig_vs_r} for simulations I and J from \cite{Dale13b}. (Run~I forms a total of 168 stars from a GMC with initial mass $10^4$M$_\odot$ and radius 10\,pc whereas Run~J forms a total of 685 stars from a GMC with initial mass $10^4$M$_\odot$ and radius 5\,pc.) 

The general increase in the fraction of triggered stars with distance from the ionising sources is a result of the redistribution of star formation discussed in \cite{Dale13b}. The ionising sources inhabit centrally--condensed clusters comprising stars which formed along with them, are therefore not triggered, and whose space--density generally declines with distance from the ionising sources. The action of feedback is to deflect potentially star--forming gas out of the clusters and sweep it into the walls of bubbles, which also gather gas from further out in the clouds that was not forming stars. Most of the star formation in the later stages of the simulation therefore occurs in this admixture of gas near the bubble walls. Most of the triggered objects are thus to be found in the bubble walls, moving away from the ionising sources, but they are mixed with a population of non--triggered objects formed from material prevented from accreting on the clusters. The triggered and spontaneously--formed stellar populations are thus well--mixed even on few--Myr timescales, and there is no appreciable gap in time between the formation of the ionising stars and the formation of the triggered population. 

The non--monotonic nature of the increase in the triggered fraction with radius owes to the clustered nature of star formation, and to the highly irregular shapes of the bubbles. This in turn is caused by the very inhomogeneous structures of the clouds and the accretion flows directed inwards toward the clusters, which results in the bubbles expanding at very different rates in different directions.

An alternative scenario considered by \cite{Dale07a}, \cite{Dale12a}, \cite{Gritschneder09}, \cite{Bisbas11}, \cite{Ngoumou15} was to consider a cloud or clump which did not possess any massive stars itself being externally illuminated by a source of ionising photons. The models of \cite{Gritschneder09}, \cite{Bisbas11} and \cite{Ngoumou15} consider the external ionisation of a stable Bonnor--Ebert sphere, and produce an effective triggered fraction of 100\,per cent, since the initial conditions are stable in the absence of feedback. This avoids the issue of triggered and spontaneously--formed populations \textit{in the target cloud} being spatially mixed. While the simulations of \cite{Dale13b} fail to generate triggered stars separated from the ionising sources by more than 10\,pc, \cite{Bisbas11} find that lower ionising fluxes (either from fainter sources, or greater separations between the sources and the clouds) result in the formation of more stars. Since the Bonnor--Ebert spheres are stable, the time interval between formation of the first stellar population and the collapse of the irradiated cloud could be arbitrarily long, but once the ionising source begins operating, evolution of the irradiated cloud is rapid. However, the stellar population which would be associated with the ionising sources in these calculations is not considered, and would in reality expand dynamically, as in the $N$--body calculations discussed above. Additionally, the clumps considered in these simulations have masses of only a few tens of solar masses and thus cannot form large clusters. Finding a smooth, stable Bonnor--Ebert sphere able to form a $\sim10^{3}$\,M$_{\odot}$ cluster in reality is unlikely.

\cite{Dale07a} and \cite{Dale12a} instead illuminated massive turbulent (and therefore intrinsically unstable) clouds, which would form stars spontaneously in the absence of external feedback. They found that the increase in the numbers or masses of stars formed due to feedback was modest, particularly if the irradiated cloud was gravitationally bound. They consequently observed that the triggered populations were contaminated by the presence of spontaneously--formed stars. In these calculations, the instability of the clouds precludes long time intervals before the initiation of star formation inside them. The separations between the photon sources and the clouds considered were only a few to ten pc. Increasing this distance is likely to reduce further the influence of photoionisation on the star formation process in these clouds since the perturbing effect on the gravitational collapse already in progress will be lessened. We note that increasing the separation between the ionising source and the target cloud to 40\,pc, as would be required to prevent dynamical mixing, entails a decrease in the ionising flux felt by the cloud by factors of order ten compared to the models of \cite{Dale07a} and \cite{Dale12a}. It is therefore difficult to imagine that the target cloud be placed far enough from the ionising source that the stellar population associated with the massive star does not become dynamically mixed with the stars formed by the cloud, yet still be close enough for the ionising flux to trigger significant additional star formation.

In principle, the spontaneous formation of stars in a GMC that has already been polluted by supernovae ejecta can deliver $^{26}$Al and $^{60}$Fe to the Solar System. However, one would expect low-mass stars from the first epoch of star formation to be present, and therefore some degree of observable age spread to also occur. This places a constraint on the number fraction of second or third generation stars that can form spontaneously in the parent cloud of the first star formation event, in that the number of second/third generation stars must be so small as to go unnoticed. 

Photoionisation is of course not the only possible triggering agent. Winds might also be invoked, and we may compare the pressure 
\begin{equation}
P_{\rm wind} = \frac{\dot{M}v_{\infty}}{4\pi D^{2}}
\end{equation}
 exerted by the momentum flux of a wind source with mass loss rate $\dot{M}$ and terminal velocity $v_{\infty}$ at a distance $D$ with the internal  thermal pressure 
\begin{equation}
P_{\rm therm} = nk_{\rm B}T
\end{equation}
 and ram pressure 
\begin{equation}
P_{\rm ram} = \mu m_{\rm H}nv_{\rm RMS}^{2} 
\end{equation}
of a molecular cloud of mean number density $n$, temperature $T$, molecular weight $\mu$ and turbulent velocity $v_{\rm RMS}$. If we take as typical values $\dot{M}=10^{-5}$\,M$_{\odot}$, $v_{\infty}=3\times10^{3}$\,km\,s$^{-1}$, $n=10^{2}$\,cm$^{-3}$, $T=100$\,K, $\mu=2.36$, $v_{\rm RMS}=3$\,km\,s$^{-1}$, we obtain for the wind, thermal and turbulent pressures $1\times10^{-12}$, $1\times10^{-12}$ and $4\times10^{-11}$\,dyne\,cm$^{-2}$ respectively. We see therefore that the wind pressure is comparable to the thermal pressure but substantially less than the turbulent ram pressure and thus unlikely to significantly influence the cloud, unless a concentration of many tens of massive O-type stars is invoked, contrary to the requirement that the total stellar population should consist of a few thousand stars.

Supernovae are the most powerful form of stellar feedback at large scales, and the interaction of single supernovae with molecular clouds has recently been modelled by \cite{Iffrig15}. Supernovae were detonated inside, on the edge of, and outside a 10$^{4}$M$_{\odot}$ turbulent cloud. Neither of the latter two simulations had any significant effect on the quantity of dense star--forming material in the clouds on 5\,Myr timescales. The supernovae failed to trigger any additional star formation for similar reasons to those preventing photoionisation triggering star formation in the simulations of \cite{Dale12a}, namely that external feedback was unable to penetrate the dense inner regions of the clouds.

\section{Conclusions}
\label{conclude}

We have analysed $N$-body and SPH simulations of star formation to investigate the probability that the Solar System could have formed in a sequentially triggered series of star formation events in order to explain the levels of $^{26}$Al and $^{60}$Fe in the protosolar nebula. This mechanism has recently been suggested as a more astrophysically likely alternative to the direct pollution of a protoplanetary disc from supernova ejecta. Our main results can be summarised as follows:

(i) If a series of star formation events are triggered by supernovae and other feedback mechanisms, the dynamical evolution driven by two-body relaxation in the first event(s) would likely result in a significant age spread, or even age bimodality in the star formation region. Determining the ages of pre-main sequence stars is notoriously difficult, and the postulated presence (or not) of age spreads of a few Myr is currently a contentious issue in the literature \citep{Soderblom14}.  We suggest that if sequential triggering were a common outcome in star formation, then age spreads (or even bi-modality) of more than 10\,Myr would be common in the local Universe. Since such age spreads are not observed in clusters, our results could be construed as part of a more general argument that sequential triggering of star formation is rare, since it is difficult to prevent the distinct stellar populations from mixing dynamically. (The lack of observed age spreads also constrains the fraction of stars that can form spontaneously at later epochs.) We will develop this idea further in later papers.

(ii) Whilst triggering can result in significant new populations of stars (in terms of number of stars and their fraction of the total mass) in SPH simulations, these triggered stars are distributed in the same temporal and spatial phase space as the first generation of stars, hence ruling out the formation of discrete  populations by triggering, as required in some models for Solar System formation. We also note that the simulations in question did not include magnetic fields. \cite{Krumholz07} showed that the presence of a magnetic field softens the shock around an expanding H{\small II} region, and is therefore likely to suppress the triggering of star formation below the levels observed by \cite{Dale13b}.

(iii) In theory, it is possible to trigger the formation of stars in a stable Bonnor--Ebert sphere, which could account for temporally separated star-formation events. However, these events are likely to be low-mass and therefore not contain the Wolf--Rayet star(s) required for $^{26}$Al pollution. Furthermore, the problem of containing the stars in spatially discrete areas of the star-forming region would still be an issue.

Our results suggest that the astrophysical conditions required for sequential triggering of two or three spatially and temporally discrete star formation events are at best highly unlikely, and that the delivery of $^{26}$Al and $^{60}$Fe to the early Solar System from this scenario is no less improbable than in the disc pollution scenario. However, we caution that outstanding issues also exist for the disc pollution scenario \citep{Gounelle08,Parker14a}, and we will investigate these further in a future paper.

\section*{Acknowledgments}

We thank the referee, Fred Adams, for his prompt and helpful review. RJP acknowledges support from the Royal Astronomical Society in the form of a research fellowship. This research was also supported by the DFG cluster of excellence `Origin and Structure of the Universe' (JED). 

\bibliography{general_ref}

\label{lastpage}

\end{document}